\documentclass[acmsmall]{acmart}

\AtBeginDocument{%
  \providecommand\BibTeX{{%
    \normalfont B\kern-0.5em{\scshape i\kern-0.25em b}\kern-0.8em\TeX}}}

\setcopyright{acmcopyright}
\copyrightyear{2022}
\acmYear{2022}
\acmDOI{XX.XXXX/XXXXXXX.XXXXXXX}

\acmConference[Proc. ACM Hum.-Comput. Interact.]{Proceedings of the ACM on Human-Computer Interaction, CSCW}{2022}{November 12--16, 2022, Taipei, Taiwan}
\acmBooktitle{Proceedings of the ACM on Human-Computer Interaction, CSCW,
  November 12--16, Taipei, Taiwan}
\acmPrice{}
\acmISBN{XXX-X-XXXX-XXXX-X/XX/XX}

\begin{document}

\title[Not Another School Resource Map]{Not Another School Resource Map: Meeting Underserved Families' Information Needs Requires Trusting Relationships and Personalized Care}

\author{Samantha Robertson}
\affiliation{%
  \institution{University of California, Berkeley}
  \city{Berkeley}
  \state{California}
  \country{U.S.A}
}
\email{samantha\_robertson@berkeley.edu}

\author{Tonya Nguyen}
\affiliation{%
  \institution{University of California, Berkeley}
  \city{Berkeley}
  \state{California}
  \country{U.S.A}
}
\email{tonyanguyen@berkeley.edu}

\author{Niloufar Salehi}
\affiliation{%
  \institution{University of California, Berkeley}
  \city{Berkeley}
  \state{California}
  \country{U.S.A}
}
\email{nsalehi@berkeley.edu}

\renewcommand{\shortauthors}{Robertson, Nguyen, and Salehi}

\begin{abstract}
  Public school districts across the United States have implemented school choice systems that have the potential to improve underserved students' access to educational opportunities. However, research has shown that learning about and applying for schools can be extremely time-consuming and expensive, making it difficult for these systems to create more equitable access to resources in practice. A common factor surfaced in prior work is unequal access to information about the schools and enrollment process. In response, governments and non-profits have invested in providing more information about schools to parents, for instance, through detailed online dashboards. However, we know little about what information is actually useful for historically marginalized and underserved families. We conducted interviews with 10 low-income families and families of color to learn about the challenges they faced navigating an online school choice and enrollment system. We complement this data with four interviews with people who have supported families through the enrollment process in a wide range of roles, from school principal to non-profit staff (``parent advocates''). Our findings highlight the value of personalized support and trusting relationships to delivering relevant and helpful information. We contrast this against online information resources and dashboards, which tend to be impersonal, target a broad audience, and make strong assumptions about what parents should look for in a school without sensitivity to families' varying circumstances. We advocate for an assets-based design approach to information support in public school enrollment, which would ask how we can support the local, one-on-one support that community members already provide.
\end{abstract}



\maketitle

\section{Introduction}

Technology increasingly mediates low-resourced and marginalized people's access to social and economic resources, like employment \cite{dillahunt2019dreamgigs, ogbonnaya2019towards, irani2013turkopticon, salehi2015dynamo}, transportation \cite{dillahunt2017uncovering}, healthcare \cite{harrington2019engaging, saksono2018family, harrington2018designing, grimes2008eatwell}, political power \cite{erete2017empowered, harding2015civic, dickinson2019cavalry}, and education \cite{strohmayer2015exploring, pinkard2017digital, thomas2017exploring}. Research has found that factors like financial cost, digital literacy, and trust in online platforms make it difficult to build technology that effectively promotes more equitable access to resources \cite{dillahunt2018designing}. Another such factor is access to information: people need to be aware of the resources and services that are available to them, and understand how to access them. In this research, we study how low-resourced families access information about public schools and enrollment in a U.S. public school district.

In the United States, low-income students and students of color face systemic barriers to educational resources. In many cities across the U.S., neighborhoods are heavily segregated based on race and income, which leads to educational segregation when students go to their neighborhood school. In response to this problem, a growing number of public school districts have opened up public schools to the entire district and have implemented policies that allow students to apply to whichever public schools they want to attend. Although these systems have the potential to provide lower-resourced students with greater access to high quality educational opportunities, school districts have run into challenges realizing this in practice. In a number of districts including New York City, San Francisco, and Oakland, schools remain segregated and unequal, even with concerted effort and resources committed to maintaining and improving enrollment systems \cite{cassano2019nyc, smith2015sf, gormley2020problem}. Research has shown that one source of these challenges is unequal access to information about the available schools due to time and resource constraints \cite{chi21, hastings2007information, andrabi2017report, luflade2020value}. In response, governments and non-profit organizations have developed new sociotechnical infrastructure to provide information to families, particularly through online information and data dashboards.

In this research, we seek to understand whether and how these kinds of interventions benefit low-resourced families. We studied how families navigate the public school enrollment system in Oakland, California. Oakland Unified School District district serves over 50,000 students, of whom 90\% are students of color, and 72\% are from low-income families \cite{ousdStats}. We conducted 14 semi-structured interviews, 10 with parents of color and low-income parents in the district, and 4 with people who work at the school district, individual schools, or local non-profits, and who have supported parents during the enrollment process (``parent advocates''). By bringing together parents' and advocates' perspectives, we gained insight into families' needs and goals, as well as how community members currently support them towards those goals.

We found that the parents in our sample were seeking a high quality education for their child, but faced challenges in reaching this goal, including finding  useful and relevant information about the schools available, and making choices between schools that balanced access to well-resourced schools against other considerations like safety, inclusion, and convenience. Consistent with prior work on information support in education \cite{villacres2019patchwork, Villacres_2020}, we found that parent advocates built trusting relationships with parents to provide personalized support. To provide relevant information, advocates leveraged their deep knowledge of the school system, as well as their familiarity with the challenging circumstances families were facing. In some cases, this included connecting enrollment support to other resources not typically considered in conversations about enrollment, such as food, housing, and healthcare.  To support families choosing between schools, advocates avoided overly narrow assumptions about what makes a school a good fit for a family, and pushed for longer-term social change outside of school choice so that every school offers the education and resources its students need.

Our findings underscore the importance of recognizing the work that already happens in communities to support people's access to information and resources in order to understand where technology could play a useful role. We advocate for an assets-based design approach to information support, prioritizing interventions that amplify and utilize assets already present in the community \cite{Villacres_2020, pei2019assets}. Viewing parent advocates' practices and relationships as assets, an assets-based design approach would ask not how we can design interventions that reduce or replace the work of advocates, but how we can \textit{support and amplify} it \cite{pei2019assets, cai2019human, irani2016hidden}. 
\section{Background: School Choice Policies and Technology}

School choice policies in the United States have been promoted on the premise that families should be able to choose a school that they believe best meets their child's needs, rather than being assigned to a school based on proximity or desegregation plans \cite{Scott2013}. In this section, we provide a brief background on this approach, then introduce the public school choice system in Oakland.

The public school system in the United States is shaped by the history of legal and de facto racial segregation in schools and housing \cite{colorOfLaw}. School choice has been closely related to desegregation efforts, first as a way for white families to avoid integration \cite{segregationAcademies}, and later as an alternative to centralized redistricting or busing plans \cite{kirp1979oakland}. These policies can take many different forms, from subsidized private school vouchers, to charter schools\footnote{Charter schools are privately run schools that receive public funding and are free to attend.}, to transfers between public schools within a district or across district boundaries. The movement towards prioritizing parents' individual choice of schools over centralized decision-making is reflective of a broader shift away from the redistributive principles of the civil-rights era towards marketization and individualism, values characteristic of neoliberalism \cite{scott2013rosa, bowe1994captured}.

Given this history, the potential for school choice to advance social and racial justice in education has always been extremely controversial.  Today, schools across the country remain heavily segregated by race and class, with large disparities in educational opportunities and resources available \cite{orfield2001, shedd2015unequal}. Proponents of school choice argue that it has the potential to improve outcomes for underserved students by allowing them to choose higher-resourced schools outside of their neighborhood, and relying on market pressures to push lower performing schools to improve \cite{hastings2007information}. Others disagree, arguing that these policies can exacerbate segregation \cite{whitehurst2017seg}, drain resources at already struggling schools \cite{ewing2018ghosts},\footnote{In the U.S., public schools receive funding per student.} and place too much responsibility on individual parents for the quality of their child's education \cite{Scott2011a}. In this work, we focus on an existing choice plan, and study how the technology designed to support this plan enables low-resourced and marginalized families' participation.

\subsection{Case Study}

In this work we study the public school enrollment system in Oakland, California. Oakland Unified School District has an intra-district ``open enrollment'' system, which means that families can apply to any public school in the district. Oakland also has more than 30 charter schools, which have a similar enrollment process to district schools. This type of school choice policy is very common in cities across the U.S., for instance, similar processes exist in New York City, Chicago, New Orleans, and San Francisco.\footnote{Allowing choice among district schools has in part been forced on school districts by a trend in federal and state policies towards increased standardized testing and reliance on mandated school transfers as an accountability mechanism \cite{hill2007nclb}. For example, under the No Child Left Behind Act of 2001, students attending schools that underperformed on standardized tests three years in a row were eligible to transfer to a higher performing school in their district \cite{hill2007nclb}.} The school district serves a student population that is 90\% students of color (48\% Latinx and 22\% African American) and 72\% low-income\footnote{Percentage of low-income students based on enrollment in the Free and Reduced Lunch program.} \cite{ousdStats}. Around 40\% of low-income students are learning English \cite{gormley2020problem}. As is the case in many other major U.S. cities, schools remain segregated and unequal across racial and socioeconomic lines. Low-income students are overrepresented at over half of the schools in the district, and the majority of Black and Latinx students attend one of these high-poverty schools. Meanwhile, several schools in wealthier neighborhoods have close to 50\% white students.

To apply for schools,\footnote{We describe the application process for district-run public schools. The application for charter schools is separate but very similar to the process for district-run schools.} families submit a ranked list of the schools they are interested in attending. The school district assigns seats using an algorithm that is designed to maximally satisfy students’ preferences, determining how to assign overdemanded seats using the school district's priorities, which include priority for siblings of current students and students who live near the school \cite{Abdulkadiroglu2019}. Students are not guaranteed admission to their preferred school, and schools give priority to students living in a surrounding zone, meaning that the most popular schools admit very few students from other neighborhoods. Recently, the school district piloted a new equity-oriented priority category at three heavily overdemanded schools, which prioritized applicants living in areas with majority low-income African American or Latinx families.

\begin{figure}
  \includegraphics[width=\textwidth]{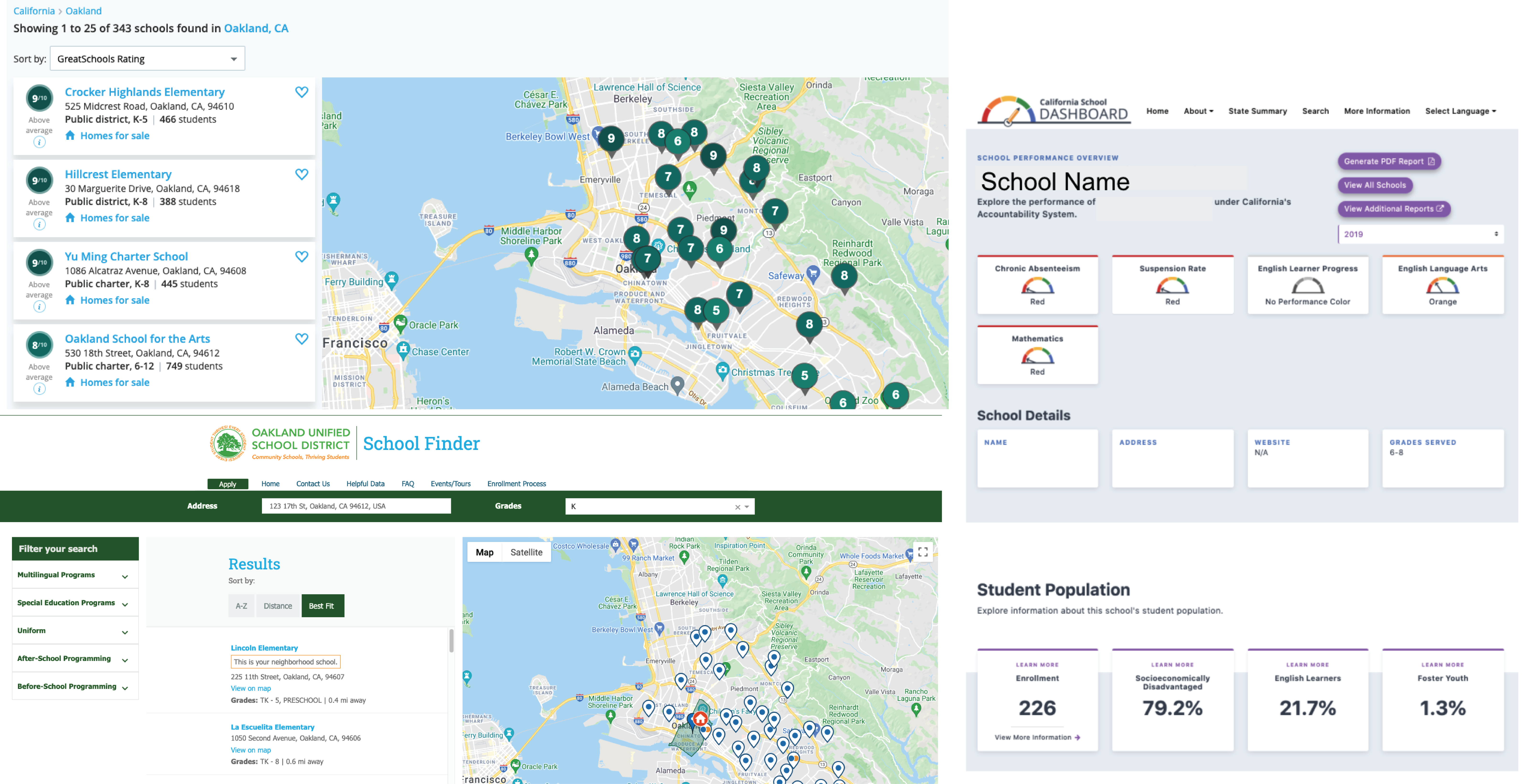}
  \caption{Three online dashboards available to families in Oakland. Top left: GreatSchools displays schools on a map, ranked based on a proprietary score out of 10. Each school has a separate page with more detailed information including the breakdown of the overall score into an academic progress score and an equity score, student demographics, and test score data disaggregated by race, income, and disability. Right: The California Schools Dashboard provides statistics on every public school in California, including standardized test scores, suspension and attendance rates, and student demographics. Bottom left: The Oakland SchoolFinder gives families the option to find their neighborhood school, filter by language programs, special education programs, uniforms, and before and after school care, and view all schools on a map.}
   \Description[Three screenshots of different school websites: \url{https://ousdapply.schoolmint.net/school-finder}, \url{https://www.greatschools.org/california/oakland/schools/}, \url{https://www.caschooldashboard.org/}]{Three screenshots of different school dashboard websites. Top left: the GreatSchools map view for Oakland (\url{https://www.greatschools.org/california/oakland/schools/}). On the left side bar is a list of schools with the address, enrollment, and a large circle with a score out of 10. In the center is a map with markers for schools that are labelled and color coded by the score out of 10. Bottom left: The map view of the Oakland School Finder (\url{https://ousdapply.schoolmint.net/school-finder}). On the left is a list filtering criteria, in the middle is a list of schools, with the user's neighborhood school listed first and the option to sort alphabetically, by distance to home, or by ``Best Fit'' (with the filtering criteria), and the right shows a map with a pin for each school and the neighborhood school highlighted. Right: The California Schools Dashboard (\url{https://www.caschooldashboard.org/}) page for a specific school, with identifying details redacted. The page shows the school name, then has a row of boxes titled, from left to right: Chronic Absenteeism, Suspension Rate, English Learner Progress, English Language Arts, and Mathematics. In each of these boxes is a gauge symbol with colors from red (left) to blue (right) and an arrow showing where the school falls based on each measure. Below this is another row titled "School Details" which contains a box for each of Name, Address, Website, and Grades Served. Finally, at the bottom is a last row titled Student Population with boxes containing the total Enrollment, and percent Socioeconomically Disadvantaged, English  Learners, and Foster Youth.}
    \label{fig:dashboard}
\end{figure}

For this system to work towards equitable outcomes, historically underserved families need to be fully informed about their options, choose to apply to higher-resourced schools than their neighborhood school, and be admitted to one of those schools \cite{hastings2007information, hill2007nclb}. The school district has invested in infrastructure to support families participating in this process. For example, the frequently asked questions page on the district's website encourages families to research the available schools using a dashboard-style school finder website, similar to that provided by GreatSchools.\footnote{https://www.greatschools.org/} The dashboard shows the available schools on a map, and then for each school includes a blurb about the school, statistics on the school's size, demographics, and academic performance, as well as a list of special programs offered, start and end times, and information about how to apply (Figure \ref{fig:dashboard}). Families can also get personalized recommendations by answering a brief survey about their child's incoming grade, home address, priorities for offerings like before or after school care, language immersion, special interest programs (e.g. STEM, arts, sports), and learning styles, and whether they care more about proximity or program offerings. Once families have narrowed down the schools that they are interested in, they are encouraged to look for more information on the school's website, or attend a school tour. Finally, families can submit an application online by submitting a ranked list of up to 6 schools. The school district also has a dedicated office that provides enrollment support in-person or over the phone. In this work, we seek to explore how this socio-technical system facilitates or inhibits low-resource families' access to educational resources.

\section{Related Work}

By studying the challenges that low-resourced and marginalized families face in navigating public school choice systems, we build on a body of work in HCI and CSCW that aims to advance social justice by broadening people's access to resources. After an overview of the broader context of HCI for social justice, we  focus on challenges and opportunities related to information access in lower-resourced communities, and how this manifests in the context of school choice.

\subsection{HCI for Social Justice}

Increasingly, HCI researchers have been interested in applying their design and research methods to address large scale social problems \cite{dombrowski2016social, baumer2011when}. This work is especially challenging because its central questions are complex, political, and often have no correct or definitive answers. Research has shown the importance of designing with the specific needs and constraints of low-resourced people in mind. For example, ridesharing technologies have great potential to actually improve people's mobility, and by extension their access to employment and other resources. However, Dillahunt et al. found that lower-resource users face significant barriers to fully realizing these benefits, such as cost, low digital literacy, and a lack of trust in the platform \cite{dillahunt2017uncovering}. Similarly, prior work has shown that technologies to support job seekers often fail to serve people with limited resources or education \cite{dillahunt2018designing, ogbonnaya2019towards}.

Even technologies that were intentionally designed to improve people's social, economic, or political position can fall short of this goal. Prior work has explored how technology can support crowd workers to increase their income and improve their working conditions \cite{irani2013turkopticon, salehi2015dynamo}, help low-income people and immigrants share information and emotional support \cite{hsiao2018immigrant, israni2021library}, and offer avenues to broaden civic engagement and participation in local politics \cite{erete2017empowered, harding2015civic}. One common theme in this work is that technological systems are limited in their ability to foster trust, build community, and shift power relations, all of which are critical to lasting social change. For instance, Erete and Burrell found that communities used a variety of technologies to engage in local politics, but existing structural inequalities shaped the impact of their practices \cite{erete2017empowered}. The authors found that wealthier, white communities were more likely to have their voices heard by local politicians than communities of color, proving that technology alone cannot equitably empower citizens. We build on this work by studying how lower-resourced families use a system that was intended to support their access to educational resources.

\subsection{Unequal Access to Information}

Information costs are one central barrier that low-resourced people face when accessing resources and support \cite{israni2021library}. A large body of work in HCI has studied how people use technologies to search and navigate information (e.g., see \cite{ko2021seeking} for an overview). However, research with lower-resourced people has shown that  making information more easily available is not necessarily sufficient to support their information-seeking needs. For instance, Israni et al. found that social norms and a lack of institutional and interpersonal trust can deter low-income people from participating in an online social network to find and share information, even when that network was a closed community of other low-income people run by a trusted organization \cite{israni2021library}. Their findings were in line with prior theories of information sharing in marginalized communities, which suggest that low-resourced people prefer to seek information and resources from trusted, close relationships, such as family or friends, because these people are more likely to understand their context and less likely to judge or patronize them \cite{chatman1996infopoor, keller2000attitudes}. 

As digital technologies like parent portals and free online learning resources have become more popular in education, research has studied how these tools impact lower resourced parents' engagement in their children's education \cite{DiSalvo2016, Villacres_2020, villacres2019patchwork, madaio2019salt, villacres2017design, cho2019comadre, khanipour2014exploring}. For example, DiSalvo et al. found that the increasing prominence of online educational tools and social networks has increased inequality between children of lower and higher socio-economic status, due to differences in information seeking practices \cite{DiSalvo2016}. One potential source of these differences is that lower-resourced and marginalized people employ culturally-specific strategies that may not align with dominant approaches to improving educational outcomes \cite{Villacres_2020}. Indeed, schools often standardize communication and materials to project a sense of equality \cite{villacres2019patchwork}, but the same resources do not benefit families equally based on their social, economic, and cultural position \cite{Villacres_2020, DiSalvo2016}. Wong-Villacres et al. explored how parent liaisons carefully craft patchworks of information to include and engage Latinx immigrant families in U.S. schools, leveraging their bi-cultural knowledge and social networks \cite{villacres2019patchwork}. We build on this work by exploring how these dynamics play a role in public school enrollment, and how centering the work of parent liaisons (or advocates in this work) can contribute to more effective support for low-resourced and marginalized families.

\subsubsection{Information and School Choice}\label{rw:schoolinfo}

There is growing recognition that participating in school choice is costly for families, who may face an overwhelming number of choices, often with little prior knowledge of the available schools or what they should be looking for in a school \cite{chen2020information, corcoran2018leveling}. A major concern with these systems is that information barriers may disproportionately impact lower-resource families, thus limiting the potential for choice policies to improve educational outcomes or promote integrated schools \cite{corcoran2018leveling, Burgess2009parental}. 

Researchers, primarily in economics, have thus been interested in how information interventions could influence families' applications to schools. Several experimental studies have found that when low-income and non-English speaking families are provided with more information about schools with high test scores or graduation rates, they are more likely to apply to and enroll in one of those schools \cite{andrabi2017report, corcoran2018leveling, luflade2020value, hastings2007information, chen2020information, bergman2020}. This kind of data about schools is increasingly available to parents, as government and non-profit organizations have produced tools to make school accountability data available on a larger scale, e.g., GreatSchools.org\footnote{https://www.greatschools.org/} or the California Schools Dashboard\footnote{https://www.caschooldashboard.org/} (Figure \ref{fig:dashboard}). However, it is not clear whether these tools can actually broaden participation in school choice. One concern is that more well resourced families use and benefit from any available information more than lower resourced families \cite{corcoran2018leveling, schneider2002}, as has been the case with other open educational resources \cite{DiSalvo2016}. Another is that emphasizing quantitative measures, especially test scores, disadvantages schools that serve more low-income students and English learners \cite{greatschoolsBiased}. This could exacerbate segregation if more privileged families use this information to avoid these schools \cite{black1999do, bergman2020}. 

While these concerns have been particularly prominent, neither centers the needs and experiences of lower resourced families. Qualitative and quantitative evidence shows that although parents across social groups value academics when choosing schools, on average, low-income families and families of color have fewer high quality options practically available, leading them to choose apparently lower performing schools \cite{Burgess2015, Abdulkadiroglu2019}. Conflicting factors include proximity to home \cite{hastings2007information, Burgess2009parental}, representation of peers from similar ethnic or socioeconomic backgrounds \cite{Burgess2015, schneider2002}, and likelihood of admission at heavily oversubscribed schools \cite{corcoran2018leveling}. For example, Hastings and Weinstein ran a field experiment where they sent families information about schools with higher test scores than their current school, and found that families were more likely to apply to a school with higher test scores if there were options close to their home \cite{hastings2007information}. Corcoran et al. gave students at high poverty middle schools a list of nearby high schools with above median graduation rates, and included information about the students' chance of admission. While the students, on average, did not apply to schools with higher graduation rates, they did use the information to choose schools where they had a higher likelihood of admission, reducing their chances of matching to a school with a very low graduation rate \cite{corcoran2018leveling}.

Taken together, this work highlights the gaps remaining in our understanding of how lower-resourced and marginalized families navigate public school choice, and how technologies, especially relating to information about schools, could best support them. In this work we aim to further this goal by bridging economists' analyses of the role of information in school enrollment and research in HCI and CSCW on marginalized parents' information seeking.
\section{Methods}

Our goal in this research is to understand how low-resourced and marginalized families navigate the public school enrollment process. In particular, we were interested in understanding the challenges that parents face in finding information about schools, completing the application, and enrolling in a school. 

\subsection{Research Partnership}

We began this research in collaboration with staff at the school district and an organization that runs charter school enrollment. We met with these partners weekly to identify our research questions and develop a recruitment strategy over the course of several months. Our collaborators were especially interested in how their information infrastructure could better support families who are harder to reach through their existing outreach strategies (e.g., online feedback surveys, outreach to parents at school sites, or hosting in-person outreach events), specifically low-income families of color and parents/guardians with limited English proficiency.

\subsection{Data Collection}

We conducted semi-structured interviews with 10 parents and 4 parent advocates. The parent advocates were people who worked to support parents through the enrollment process as staff at a school, the school district, or a non-profit. Three of the four parent advocates had also enrolled their own children in public schools using the same enrollment system, giving them a mix of parent and advocate perspectives.  To preserve participant privacy, we identify participants in this paper by unique identifiers: P1 through P10 for parents, and A1 through A4 for parent advocates.

\subsubsection{Recruitment} In order to target lower resource families, we distributed a brief screener survey in English and Spanish that asked for the respondent's race, zipcode, the year they most recently enrolled a student in a public school, and how easy or difficult they found the enrollment process (5-point Likert scale).\footnote{Since responses were sparse, our only exclusion criteria was applied to white parents. The exception to this criteria was A1, who we recruited for his unique perspective as a school principal.} Our recruitment materials also included a phone number that prospective participants could text for more information. The charter school enrollment organization and a number of community-based non-profits shared a recruitment flyer  on our behalf through their social media accounts.  

P1, P2, and A1 were recruited through this recruitment survey, but responses to this survey were very slow, and most parents who filled it out did not respond when we reached out to set up an interview. In light of this, we turned our attention to building stronger connections with people who work with families in our target groups. We met people who worked with families in a wide variety of capacities, from people who specifically provide enrollment support, to school leadership and staff, to education-related non-profits and activist organizations. Through this process we met and interviewed A2-A4, and we had informal conversations with several other people. While our initial target sample was parents or others who participated in enrollment, we quickly realized that these community members played a key role in informational support. Thus, our analysis of families' experiences navigating this process is enriched by the perspectives of those who support them. Throughout the paper we refer to this group of participants as ``parent advocates'' for simplicity. However, while there are some commonalities across their roles and perspectives, each brings a unique perspective.

\subsubsection{Interview protocol} During the interview, we asked parents to describe their experiences applying for schools and enrolling their children. We were particularly interested in what parents wanted in a school, how they found information about schools, and their experiences submitting the application and receiving results, but we encouraged participants to speak about whatever part of the enrollment process was most salient for them. We asked parent advocates to describe their work, the challenges they face, and how they felt the enrollment system could better serve the needs of the families they work with. Interviews were between 30 and 50 minutes and participants received a \$30 gift card. Interviews were conducted over the phone in English and Spanish (with an English-Spanish bilingual interpreter) between September 2020 and March 2021.

At the end of the interview, we asked parents a set of open-ended demographic questions. Parent demographics are shown in Table \ref{demographics}. As the research proceeded we adjusted the demographic survey in an effort to respect participants' comfort level while also ensuring we reached a diverse set of lower resource participants. For this reason, the information in Table \ref{demographics} is not consistent for every participant. For instance, in earlier interviews we asked for participants' home zipcode as a proxy for income, while in later interviews we asked participants for their highest level of education, whether they were employed and, if they were employed, whether their income was above or below the city's median. Table \ref{advocateroles} shows parent advocate demographics and a description of their role working with parents.

\subsubsection{Limitations and Opportunities} Our sample in this paper is small because we spent significant time and effort recruiting participants and building relationships with community groups. Certainly, our data does not represent all of the different experiences of families of color, low-income families, or immigrant families in the school district. By interviewing parent advocates and some of the parents they have worked with, our sample represents the perspectives of families who needed more support and were able to seek that out. Our findings should not be taken to indicate that all families rely heavily on parent advocates. Despite these limitations, we believe the findings provide detailed insight into a subset of families' experiences, and are indicative of some challenges that lower-resourced and marginalized families can face. Further, our findings are consistent with prior research into marginalized families' experiences and information practices in other areas of public education \cite{villacres2019patchwork}.

Future work should engage with families from a wider range of backgrounds, including people who speak other languages, and people who did not receive enrollment support. We use the term ``parent'' throughout this paper because it accurately describes our participants, but there are other people who participate in the enrollment process, like foster guardians and older students. School districts wishing to replicate this work should define target groups of families and engage with families in each one in order to understand the range of local needs.

\begin{table*}
\caption{Parent demographics. Interviews with participants marked with $\dagger$ were conducted in Spanish and English with an interpreter. Incomes marked with * are estimates based on median income in home zip code, all other fields are self-described.}
\label{demographics}
\begin{tabular}{l l l l}
    \toprule
     & \textbf{Race/Ethnicity} & \textbf{Income} & \textbf{Education} \\
    \midrule
P1	&	African American	&	Low*	&	-		\\
P2	&	Filipino	&	Low*	&	-		\\
P3	&	African American/Black	&	Low*	&	-		\\
P4	&	-	&	-	&	- 	\\
P5$^\dagger$	&	Honduran	&	Low*	&	-		\\
P6$^\dagger$	&	Latino	&	Below median	&	Middle school 		\\
P7$^\dagger$	&	-	&	Unemployed	&	None\\
P8$^\dagger$	&	Guatemala	&	Unemployed	&	2nd grade		\\
P9 & - & - & - \\
P10 & African American & Below median & Some college\\
    \bottomrule
\end{tabular}
\end{table*}

\begin{table*}
\caption{Parent advocate roles and the target groups they serve. Race or ethnicity was self-described, the authors inferred role descriptions and target groups from the interview transcripts. Three of the four advocates have also enrolled their own children in public schools, and shared perspectives in their interview both as a parent and a parent advocate.}
\label{advocateroles}
\begin{tabular}{l l l l l}
    \toprule
     & \textbf{Race/Ethnicity} & \textbf{Parent} & \textbf{Role} & \textbf{Target groups} \\
    \midrule
A1	&	White & x	&	School principal & School is 50\% low-income students	\\
A2	&	-	& x &	Non-profit staff & Primarily Black and Latinx families	\\
A3	&	Latino	& x &	School-Parent liaison & Primarily Spanish-speaking families	\\
A4	&	White	& &	School district staff & Recently arrived refugee families	\\
    \bottomrule
\end{tabular}
\end{table*}

\subsection{Data Analysis}

We analyzed the interview transcripts using inductive, qualitative analysis \cite{qualresearch}. Working from prior literature, the first author developed a code book containing 32 codes identifying sources of information that parents use and challenges they face when enrolling their students. These codes mostly fell into three higher-level categories: information (e.g. ``online,'' ``school visits''), factors of consideration for schools (e.g. ``test scores,'' ``convenience,'' ``safety''), and broader concerns about the process (e.g. ``finding a good school,'' ``systemic racism''). Next, we conducted open coding on a line-by-line basis \cite{groundedtheory}. The first two authors worked together to code two transcripts this way, resolving disagreements through discussion. Each author then analyzed half of the remaining transcripts. The initial codebook was used as a reference, but codes were adjusted, added, and removed as necessary to best fit the data. At the end of this process, the two authors discussed their findings and again resolved disagreements. The final codebook contained 39 codes. There were two natural groupings of codes (``finding information'' and ``considering priorities''), which contained 25 of the 39 codes. Other codes included ``building relationships,'' ``worrying about availability/scarcity,'' and ``voicing concern.'' Next, we conducted axial coding to identify relationships between codes and extract higher level themes. 

Parents' experiences varied widely based on their prior experiences, access to social and economic resources, and the level of support they received from other people, and were further shaped by intersecting identities of class, race, language, and education. We do not mean to imply that ``low-resourced'' or ``marginalized'' are fixed or stable categories. To the contrary, many parents we interviewed faced a mix of experiences of privilege and marginalization at different points when participating in school choice \cite{crenshaw2017intersectionality}. We strive to avoid artificial distinctions when presenting our results, and instead aim to present an authentic account of participants' experiences, using their own words as much as possible.

\subsection{Research Approach and Researcher Positionality}

In this work we sought to engage low-resourced and marginalized parents to understand how they navigate the public school enrollment process. Prior work has discussed the challenges for researchers engaging with low-resourced populations in equitable and just partnerships. For example, Harrington et al. discussed how collaborative design workshops with underserved communities can cause harm when researchers are not sensitive to the communities' historical context and experiences of oppression \cite{harrington2019deconstructing}. In order to minimize these kinds of harms, researchers have called for new approaches and methodologies that account for historical contexts and systems of oppression (e.g. \cite{dombrowski2016social, costanza2018design, harrington2019deconstructing, irani2010postcolonial, ogbonnaya2020critical, bardzell2011feminist, liang2021embracing}). In this section, following feminist methodology, we reflect on our approach to this research and the ways in which our positions in the world, our goals, and our beliefs shape our interpretation of the findings \cite{bardzell2011feminist}.

Prior work has emphasized the importance of working with trusted community members to ensure that research with marginalized communities is conducted in culturally appropriate ways, and to gain legitimacy and trust with participants \cite{harrington2019deconstructing, costanza2018design}. The online screener survey approach was not effective, possibly because parents did not trust an impersonal online flyer, did not have the time to fill out a screener survey, or were hesitant to participate in research they felt would not serve them. We were also sensitive to the additional burden of the ongoing COVID-19 pandemic, during which many parents had increased childcare responsibilities, and which had a disproportionate and devastating impact on Black and Latinx communities in the United States \cite{lakhani2020covid, wood2020covid}. Reflecting on this process, it would have been better to begin by engaging not only with the school district, who hold significant institutional power and whose interests may not represent those of marginalized community members, but with various community leaders to develop research questions and methods. We acknowledge that had we taken this approach, this research may have looked very different in terms of central questions and methods.

None of the authors come to this work with first-hand experience of a public school choice system, neither as parents nor students. The first author is a white woman from a class privileged background and does not experience the forms of marginalization that we discuss in this work. The second author is a first-generation college student and comes from a family of Vietnamese refugees. The last author is an immigrant and woman of color who attended public school at her home country. Our lack of shared context with participants likely shaped our approach to this work, the kinds of experiences and insights that participants shared with us, as well as our interpretation of those that they did share. The authors have a range of disciplinary backgrounds, but primarily adopt an HCI lens to study a specific sociotechnical system situated within much broader conversations about education and social justice. 

Certainly, our perspectives have been influenced by working in partnership with staff members at the school district and charter school enrollment system, although we have not consciously shaped our findings in any way to meet their expectations or censor families' negative experiences. We have shared our findings back to the district partners and parent advocates to verify our interpretations and contribute to ongoing discussions regarding improvements to their enrollment process that could promote more equitable participation and access to schools.
\section{Results}

Next, we present our findings about low-resource families' information needs in the school enrollment process. We organize our findings based on the two major stages of the application process: finding information about the available schools, and forming a ranked list of schools. First, we summarize the challenges that parents faced, then we highlight parent advocates' strategies to provide support. 

\subsection{Available information lacks relevance and personalization}

When they are first participating in school choice, parents need to find out how the system works and learn about their options. Our findings indicate a mismatch between the information families need and the information that is easily available. Further, families are faced with a potentially overwhelming number of options, which exacerbates the challenge of finding relevant and helpful information. To address this challenge, parent advocates build trusting relationships with families and learn about their specific needs and circumstances. This enables them to make personalized recommendations and provide more holistic support for families. 

\subsubsection{Challenges}

Families face challenges learning about the choice process and why they may want to participate, and finding useful information relevant to their priorities. On top of this, they are faced with a potentially overwhelming number of options to learn about.

Before they even begin to evaluate different schools, families need to be aware that they have a choice to apply to different schools, and then find information about their options. If parents are not aware of the resource disparities between schools in their district, they may not be motivated to participate in school choice. As A2 told us, \textit{``you assume that people who have a degree or people who are educated are going to educate your babies.''} In fact, several parents we interviewed only learned of the choice system after they had enrolled their children in their neighborhood school. For example, P8 heard from other parents at her child's school that \textit{``there are way better schools.\ldots They say far away schools \ldots are better, but I don't know, I'm not sure about that.''}

Next, once a family has decided to participate, they need to find information about the available options. However, we found that available information often lacked relevance and utility to families. For example, school websites were generally poor quality except at high resource schools, and some parents found the information available to be tailored to the interests of higher resource parents. For those schools,

\begin{quote}
    \textit{I mean, you know everything, you got teacher bios, walkathon, the hundreds of thousands of dollars you can raise, how you can pay it, sign up for afterschool. \ldots it's geared towards parents who are looking for certain things. It's geared toward a middle-class, upper middle-class aesthetic, because the other schools, that's not where they're putting their resources, understandably. (P1)}
\end{quote}

As discussed in section \ref{rw:schoolinfo}, concerns about information access have driven the growth of dashboard-style websites like GreatSchools\footnote{https://www.greatschools.org/} or the California School Dashboard\footnote{https://www.caschooldashboard.org/}, which provide school statistics, like test scores and demographics, in a digestible format. While some parents in our sample found this kind of data useful, others pointed out that it is still limited, especially if you are concerned that your child might face more adversity than the average student. P3 tried to use test scores disaggregated by race to find a school that would be a good fit for her daughter. However, she pointed out that average scores are only rough indicators of how a school will support your child:

\begin{quote}
    \textit{[After looking at test scores] you are thinking, ``Oh, my child's here,'' and then you take the test and they're like, ``Well, actually, they're behind.'' Now, are the school going to help me get her up to what she needs to be? Do they offer or have those resources available? (P3)}
\end{quote}

The district encourages families to use online information to narrow down the schools they are interested in, and then attend in-person tours to learn more. Many participants agreed that tours provide a valuable opportunity for parents to assess whether their family will be safe and included at a school. For example, A2 tells parents to visit the schools that they are interested in to make sure that they \textit{``feel welcome.''} However, prior work has found that school tours are time-consuming and often inconvenient for parents with inflexible work or childcare schedules \cite{chi21}. Participants in this study discussed additional challenges with school tours that were specific to families of color. For P1, \textit{``it felt very isolating to be looking for schools as a middle-class Black parent in [city name].\ldots We went to the [school name] tour for my five year old, lines out the door, full cafeteria, people asking questions like, ``What's your track record on where you get kids into college?''''} A1 recalled, \textit{``my wife was the only non-white person on the tour. You know, like they had all this like Spanish translation available [but] nobody needed it.''} Despite efforts to include families of color, for example, by providing interpreters, A1 found that parents still struggled to get useful information specific to the experiences of children of color during school tours:

\begin{quote}
    \textit{People have real questions, \ldots especially around identity safety, whether or not there are other kids like theirs at the school, things like that. \ldots I went on some tours and I saw people asking questions towards that question and the tour guide either kinda didn’t pick up the gist of the question, the nuance of it, or they did and they answered it in a like not real way. (A1)}
\end{quote}

Overall, participants identified a mismatch between the information that is easily available to families, which assumes a level of understanding of how the system works and a certain set of interests, and the information that parents need, particularly regarding how lower-resourced students are included and academically supported at each school.

\subsubsection{Strategies for support}

Parent advocates provided more relevant and concise information to families by first building a trusting relationship with them, learning about their personal circumstances and priorities, then connecting them with the right information or resources. There are over 80 district schools and more than 30 charter schools in Oakland, so much of this work involved reducing information overload for families. In addition, this support often extended beyond enrollment to address other survival needs like housing, food, and healthcare.

When A2 starts working with a parent, the first step is \textit{``a conversation of getting to know the parent and what it is that they're ultimately looking for.''} This conversation is not simply eliciting a set of criteria that the parent already has in mind, but is a more personal process of \textit{``learning about the family and what it's going to take for that child to do well and be successful.''} A2 brings together parents' expertise on their family's circumstances, with her own expertise on schools and education, to recommend schools that will meet the family's needs within their constraints.

A4 works with refugee families who have recently arrived in the United States. Rather than providing detailed information about all of their school options, he accounts for these families' difficult circumstances when providing support.

\begin{quote}
   \textit{Usually when families are coming to enroll with me \ldots they're totally overwhelmed with the whole process and the information overload so the idea of like sitting there and talking through like ups and downs, all the ins and outs of each school it's like, no, they don't want that, they want a couple of options, and to feel like I am a trustworthy person \ldots that cares about their students and going to give them the right information. (A4)}
\end{quote}

He emphasized that this kind of support relies on trust and care:

\begin{quote}
    \textit{My absolute nightmare is somebody coming to me and being like, ``You know what we need is like a resource map or like a resource guide so we can start giving more information out to families.'' I'm always like that is the last thing we need, we need more people to explain things to people in person or one-on-one, and answer their questions. Because, especially if you're new here, or don't have a lot of education yourself, the idea of navigating a website and filling out forms. It's just doesn't\ldots it's\ldots it's ridiculous. So, I think that the best way is just more personalized. (A4)}
\end{quote}

In order to build trust with these families and provide relevant and helpful support, advocates considered their circumstances holistically, and where necessary, connected families to other resources. One reason that families might be looking for a new school is that they are experiencing some form of instability, for instance, they have recently immigrated and/or they have unstable housing. In these circumstances, enrolling their child in school may be only one of many challenges they are facing. For instance, when we spoke to A4 he was working with a mom who had very recently arrived in the United States. He not only helped her enroll her son in a school, but also organized for him to receive the required immunizations and signed him up for free meals through the school district. He also provided her with financial support and was helping her find free legal representation. P6 also worked with A4 to enroll her son in school. She still frequently reaches out to him for support, and he has helped her, \textit{``know about the food they give out in school, \ldots connect with the internet service and \ldots finding the router, and then he has helped me with my son's computer.'' } A3, who works at a school, also connected families with different sources of support, such as financial aid, mental health care, housing, and legal aid. 

We found that this type of support could also be helpful for other low-resource families, but that it was difficult to find. For example, P4 was frustrated by a lack of support from her son's school to find permanent housing.

\begin{quote}
   \textit{I was searching for honestly like resources for housing because like when I enrolled him they were saying how if you need any assistance in anything [they could help.] \ldots They did sign him up for free meals to be delivered, so that was one thing they helped with. \ldots But as far as anything else it's not really a great experience honestly. (P4)}
\end{quote}

Typically, discussions of information support for school choice focus on information about schools. This assumes that families can afford to spend significant time and effort researching different school options, and overlooks other elements of the process, like ensuring students have the immunizations and technology they need to participate in school activities, as well as the food and housing security they need to focus on learning. Parent advocates' strategies focused on personal relationship building and trust and highlighted the importance of attending to individual families' circumstances and priorities.

\subsection{Ranking schools involves difficult trade-offs}

Families apply for schools by submitting a list of schools ranked in the order of their preference. We found this process can be challenging for families, even aside from the challenges of finding relevant information outlined above, because they need to know what to look for in a school, and often face difficult trade-offs in deciding which schools would be the best fit for their family. To support families making these decisions, advocates avoid overly narrow assumptions about what makes a school a good fit for a family, and push for longer-term change so that every school offers the education and resources its students need.

\subsubsection{Challenges}

Parents may not know what to look for in a school, and even those who both know what they want and have relevant information available may face complex trade-offs between those priorities.

School choice has created an unusual situation in which parents are expected to have substantial knowledge about what makes for a quality education. Some participants, particularly those who had recently migrated to the U.S. and/or who had fewer years of formal education themselves, felt unsure about what to look for in a school. For example, when we asked P6 to describe the ideal school for her son she told us, \textit{``I wouldn't know because I don't really know a lot of schools.''} Some parents found it easy to pick their top choice school, as they had specific criteria or personal connections, for instance, P6-P8 prioritized proximity to home, and P5 wanted to enroll her children in the school where their former teacher had recently transferred. However, many of them did not have back-up options in case they were not assigned to their top choice.

By claiming that offering students the choice to leave their neighborhood school increases educational equity, open enrollment policies can give the impression that some schools are desirable, while others are not. As A1 put it,  \textit{``Why would we have school choice, unless we are saying that there are some schools that are not good enough for your kids?''} However, even with choice, higher resource schools are not practically available to every family. Due to a long history of systemic racism in education and housing, higher resourced public schools remain concentrated in affluent neighborhoods and serve more white students. As a result, low-income families and families of color who invest in school choice face complicated trade-offs to find a school that provides resources (e.g. experienced teachers, high quality equipment and facilities, and smaller class sizes) but also offers safety, inclusion, and convenience. 

For example, P1 repeatedly described her experience of finding a school for her children as a \textit{``balancing act.''} For P1, the most important factor about a school was \textit{``achievement, but,}

\begin{quote}
    \textit{I wouldn't just say test scores, \ldots I also was looking at resources, \ldots I wanted a school where I felt like, as an African American family, or as a Black kid, my child was not going to be the only [one]. And I also wanted a school where there was going to be some socioeconomic diversity, and where I was going to fit in with the parents. And so, it was a balancing act of wanting to be at a place where I felt like... because for Black kids, there's so much literature, there may be a school that's a good school, but that doesn't mean it's the best school for my kid. (P1)}
\end{quote}

Economists have sought to reduce information overload by prioritizing information about schools that perform well on one or more quantitative measures of school quality, such as test scores, chronic absenteeism, or graduation rates \cite{andrabi2017report, corcoran2018leveling, hastings2007information}. However, this approach makes strong assumptions about which schools families \textit{should} prefer, overlooking these kinds of trade-offs as well as other factors that contribute to a student's experience at a school.

A1 is a principal at an elementary school where around half the students are from low-income families. He didn't believe that high resource schools are unequivocally better for students from low-resource backgrounds:

\begin{quote}
    \textit{I don’t think they really know what it means to support a child that's, you know, catching the bus from [low-income neighborhoods] and has experienced severe trauma, you know, maybe has an unstable living situation, stuff like that. \ldots There’s so much growth that has to happen in that community before that kid even really has a chance at a school like that. (A1)}
\end{quote}

A4 encourages newcomer refugee families to attend schools where there is an existing community from the same country, so that students have \textit{``their community members there already and \ldots the teachers already have done professional development to learn the background of students from these places.''}

P1 pointed out that political power dynamics at the school district level are another factor to consider. After her son had \textit{``a really rough experience ... with a teacher who just really was very anti-Black and racist,''} she was forced to move him to another school. This time she chose a school in a wealthier neighborhood, despite it having fewer Black families, partly because she felt that \textit{``the resources and the privileges there are going to mean that when there's a problem that [the school district] is going to listen.''} Unfortunately, three other parents (P3, P10, A3) in our sample reported similar stories where their child was in an unsafe classroom environment due to treatment by an elementary school teacher.

Families can rank up to six schools on their application, and are strongly encouraged to submit complete applications to maximize their chances of being assigned to at least one of their preferred schools. Some parents have a specific school in mind, for instance, the one closest to their home, in which case ranking schools is easier, as long as that school is not oversubscribed. However, many families face difficult trade-offs between priorities like academics, convenience, and inclusion, induced by the inequitable underlying distribution of resources across the city. 

\subsubsection{Strategies for support}

Advocates considered school quality broadly and accounted for families' circumstances when making recommendations or helping them navigate trade-offs. Most importantly, several of the people we interviewed viewed their work as a fight for long-term social change rather than exclusively enrollment support, and sought to ensure students had the resources they needed no matter which school they ended up at.

The level of information advocates provided to families depended on their circumstances and priorities. For example, A2 narrows the set of choices and recommends schools for families based on how they've described their priorities, but always leaves the final decision to the family.

\begin{quote}
    \textit{We just kind of let them know, ``Hey, out of the schools, based on the information that you gave me, here are some of the schools that when we look at data and information, these are like your top eight schools,'' and we always give them at least five to be able to choose from. [\ldots] We never tell the person, ``this, this, this, this, this.'' You rank it how, you know, works for your family.} (A2)
\end{quote}

A4 made stronger recommendations to families, but was sensitive to their constraints and helped them work within them. For example, sometimes the school with the strongest existing community of immigrants from a newcomer's country of origin is further from home. Proximity is often the top priority for these families (P5, P6, P7, P8), so he works outside of the formal enrollment system to arrange free bus tickets for the students to ease this trade-off.

Ultimately, school choice policies can only address inequalities in resources across schools by providing students from underserved communities the opportunity to access higher resourced schools. There are a limited number of seats available at higher resourced schools, so there will always be some students who cannot access those schools, even if they apply. 

\begin{quote}
    \textit{``There's not a whole lot of [\ldots] quality schools that are in the city, [\ldots] and the ones that are high demand, quality schools, everybody else know about them, too. So what are my chances of being able to actually get into those schools?'' (A2)} 
\end{quote}

This means that supporting families to participate in the choice system is a limited and insufficient avenue to promote longer-term justice in education. Unlike resources focused solely on choice, parent advocates were able to balance between helping families access resources within the existing system while also pushing for a future in which the choice system isn't necessary. For example, part of A2's work involves advocacy at the district-level, as well as grassroots organizing with parents. Another strategy is targeting resources and support to students who enroll in lower resourced schools. For instance, at a school where several families that A4 worked with attend, they were able to,

\begin{quote}
    \textit{``set up English classes for parents at that site and then summer school classes for students and concurrently ESL [English as a second language] classes for the parents. And [\ldots] when community-based organizations were looking for office space at that time we managed to get them space in the school and in offices close to that school.'' (A4)}
\end{quote}

In contrast, broader discussions of school choice and research into low-resource families' preferences for schools too often assume that the best way to move towards a more equitable education system is to nudge these families towards historically higher resource schools, denying the complexity of the landscape that they are forced to navigate and the limits of individual choice for promoting long-term justice.

\section{Discussion}

As school choice has expanded as a way to increase access to high quality public education, so has the concern that these systems require too much time and effort from families. Time and resource constraints disproportionately exclude underserved families, who were positioned as the main beneficiaries of choice to begin with. In an effort to reduce these barriers to participation, researchers, governments and non-profits have developed online tools and informational resources that rank and sort schools. The broad success of sites like GreatSchools indicates that these kinds of technologies are helpful for many families. However, an abundance of examples within and beyond HCI research have shown that even technology that is intended to improve underserved people's access to information and resources can struggle to do so in practice \cite{salehi2015dynamo, hsiao2018immigrant, israni2021library, erete2017empowered}. Our goal in this work was to understand the extent to which a primarily online enrollment system supported lower-resourced families participating in open enrollment, with a focus on technologies that provide information about schools and how to apply. Our findings provide insight into the information mismatches and difficult trade-offs that families face in this process. Consistent with prior work \cite{villacres2019patchwork}, our findings highlight how parent advocates provide personalized support to fill the gaps in the information infrastructure. In this section, we discuss the implications of our findings for online informational resources and their potential to address participation barriers in public school choice. Then, we discuss paths forwards, advocating for an assets-based approach that recognizes and resources the work of community members that already goes on to support families and promote longer-term social justice.

\subsection{The limits of informational resources for promoting educational equity}

We began this research hoping to identify ways in which the online enrollment system could be improved, for instance, what kind of information dashboards should contain in order to be relevant to low-resourced families. However, our findings highlight three specific limitations of improving informational technologies for promoting more equitable participation in enrollment:

\begin{itemize}
    \item First, we found that \textit{web-based informational resources lack personalization and nuance}, which are invaluable for lower resourced and marginalized families. How could an online dashboard tell a family whether they will feel safe, welcomed, and heard in a school community? Providing relevant, helpful information requires understanding a family's circumstances and goals, as well as nuanced and detailed information about the community and resources at each school.
    \item Second, \textit{prioritizing information about schools assumes that families} should \textit{gather as much information about schools as they can.} However, for some families it is entirely reasonable not to commit time and resources to searching for a school \cite{Scott2011a}. Some families are not aware that they need to apply for schools, and thus do not even seek out information to begin with. Even those who are aware of their options may not know what they should look for in a school. Our interviews highlighted that parents are navigating an inequitable system, in which they have a low probability of admission at higher resourced schools, and where those schools, if they are admitted, may expose their children to isolation and racism.  Meanwhile, they may be facing much more urgent survival needs. Providing relevant and useful support in some cases may involve deprioritizing information about schools and connecting families to other critical resources like food, housing, or healthcare.
    \item Third, \textit{online informational resources are generally targeted to individual families, which does not build community or trusting relationships, or promote longer-term social change.} Being in community with other people experiencing similar challenges creates opportunities for families to form a shared understanding of those challenges and find ways to exert collective political power to make change \cite{spade2015normal}. Further, technologies designed for individuals can overlook the social, economic, and political context that produces an inequitable distribution of needs and burdens in the first place \cite{dombrowski2016social}.
\end{itemize} 

Our findings in this research remind us to remain wary of purported tech ``fixes'' to long-standing social problems \cite{benjamin2019race, toyama2015geek, baumer2011when}, and to think critically and reflexively about how the methods and approaches of HCI and computing more broadly can contribute to social change \cite{abebe2020roles, veale2018fairness, dombrowski2016social}. Ultimately, new or improved tools and online resources focused on enrollment would benefit some families, but it would also create more work for parent advocates to support the lowest-resource families in accessing and utilizing these tools. New technology rarely \textit{reduces} the overall amount of work that needs to happen to make systems run smoothly, rather it displaces that work, often onto the most marginalized and in ways that are invisibilized and under-resourced \cite{suchman1995making, bowker2016layers}. 

Certainly, some technical improvements to the system may still prove worthwhile. In fact, many parents we spoke to strongly appreciated the district's introduction of an online enrollment option, including two parents who did not have stable housing and were able to complete the entire process on their mobile phone. However, we should not expect new technology to easily solve participation barriers in enrollment, like a lack of time to invest in the process, or conflicting factors like proximity or inclusion barring access to higher resourced schools. The cost of the substantial labor and resources needed in order for new technology to be inclusive should be factored into decisions about what new technologies to pursue and where to invest resources. To conclude, we advocate for an assets-based design approach to guide the development of interventions that support low-resourced families.

\subsection{An assets-based approach to enrollment support}

Assets-based design is an approach to designing interventions that are sustainable and useful to communities in the long-term, especially in low-resourced settings \cite{pei2019assets, Villacres_2020, cho2019comadre, mathie2003from, kretzmann1996assets, brooks2013making}. This approach is grounded in a deep understanding of people's capacities and assets, rather than looking to solve their needs or deficits \cite{kretzmann1996assets}, and has an established history in education research \cite{moll1992funds, Villacres_2020, yosso2005whose, harper2010antideficit}. One way of doing this in HCI is to design interventions that leverage and amplify existing resources and practices in a community and minimize technical novelty \cite{pei2019assets}. Allowing parents to enroll over the phone or a mobile-friendly website, for instance, did not require sophisticated technical innovation, but allowed many of the parents in our sample to participate using a device they had easily available and practices they were already familiar with. In this work, parent advocates offered rich insight into their successful practices for supporting families through enrollment both within and outside of the constraints of the existing system. Advocates did not simply nudge parents to access existing information and resources, but augmented those resources with personalized informational and emotional support, and worked in partnership with families to meet their specific needs. Viewing these relationships and practices as an asset in the community, an assets-based design approach would ask not how we can design interventions that reduce or replace the work of advocates, but how we can \textit{support and amplify} it \cite{pei2019assets, cai2019human, irani2016hidden}.

A direct way to support the work of parent advocates is to provide resources for this type of work, e.g. by hiring more people who are trained to provide personalized, culturally relevant, long-term support to more families. Many parents in our sample worked with a parent advocate because of our snowball sampling approach, but in general most people do not have access to someone with both deep knowledge of the school system and the time to provide personalized support. Although the district's enrollment center provides support in person and over the phone, staff at this center serve the entire district, and thus may lack the time and specialized knowledge (e.g. about resources specific to newcomer refugees) to provide in-depth and personalized support. In addition to district staff, it may also be helpful to train trusted community members, e.g. at community centers or places of worship, to either help families directly or get them in contact with a more specialized parent advocate.
 
A complementary approach could be to build systems that support parent advocates' work. Identifying such opportunities was not our focus in this work, but our data points to some promising directions for further investigation. For example, community advocates in the district recently showed that the enrollment system admitted very few non-neighborhood students to popular, high-resource schools, and contributed to a new priority system that will make more space for low-income students of color at those schools. One way to support this work could be to design more transparent systems that make it easier for stakeholders to identify harmful aspects of the system's design and suggest improvements \cite{veale2018fairness}. Another challenge that advocates in our sample faced was connecting families to resources beyond enrollment, like mental health care and housing. This suggests the potential for tools that help advocates build, maintain, and share networks of support and resources in their area. This echoes similar findings in other domains, for instance, environmental justice \cite{aoki2009vehicle}.

Finally, while support during enrollment can be useful, it is important to be wary of interventions that are too narrowly centered on school choice and enrollment. Advocates and many parents agreed that choice is a limited avenue to justice, and the top priority should be ensuring that every school is high quality and well resourced. For instance, while admitting more low-income students to historically overserved schools is an improvement, there are still a limited number of seats at those schools and many more students who will not be admitted. While advocates for school choice position it as the best way to ``empower'' parents and promote equity in education, it is far from the only option. In fact, Scott points out that the individualistic, market-based principles underlying school choice are not only in tension, but in direct conflict, with the redistributive principles that defined 20th century civil rights movements and continue to mobilize grassroots movements for educational equity \cite{scott2013rosa}. Across the country, students, parents, and teachers have organized for smaller class sizes, experienced teachers, community engagement in school governance, and more equitable funding structures, in an effort to guarantee a high quality education for historically underserved students without forcing them to move schools \cite{integratenyc, CTUstrike2012, schoolcolors, scott2013rosa}. Our participants surfaced tension between using the existing enrollment process to access resources \textit{today}, and working towards a more just arrangement in the future. In considering how technologies could support families and advocates, we should design to surface, rather than obscure, these tensions and avoid entrenching the dominance of individual choice over grassroots strategies to promote educational justice.

\section{Conclusion}
In this work, we found that low-resourced families and families of color faced challenges navigating a public school enrollment system that was intended to improve their access to educational resources. In particular, the information provided about schools often lacked relevance and did not account for the difficult trade-offs that families must navigate when choosing between schools. We found that parent advocates provide personalized, community-based support that cannot be emulated by impersonal and generic online resources. We reflect on this research as an example of the importance of an assets-based design approach to supporting communities' ongoing work to advance social justice.

\begin{acks}
We sincerely thank our participants for sharing their experiences and insights with us. We also thank staff at the school district and charter enrollment organization for their collaboration developing this research. We are immensely grateful to Moonhawk Kim for his support in all aspects of this research, especially making connections with local community members. We thank Lucinda Matias Mendoza for her work connecting us with Spanish speakers and helping us ensure our research practices were culturally relevant and appropriate. This work was supported by the National Science Foundation award IIS-2131519.
\end{acks}

\bibliographystyle{ACM-Reference-Format}
\bibliography{refs}

% \appendix

\end{document}